# A STUDY OF THE ENERGY DEPENDENCE OF RADIATION DAMAGE IN SUPERCONDUCTING COILS FOR A NEXT GENERATION MU2E AT PIP-II[*][†]

V. Pronskikh[‡], D. Glenzinski, K. Knoepfel, N. Mokhov, R. Tschirhart

Fermi National Accelerator Laboratory

The Mu2e experiment at Fermilab is being designed to study the coherent neutrino-less conversion of a negative muon into an electron in the field of a nucleus. This process has an extremely low probability in the Standard Model, and its observation would provide unambiguous evidence for beyond the standard model physics. The Mu2e design aims to reach a single-event-sensitivity of about 2.5 x $10^{-17}$ and will probe effective new physics mass scales in the $10^3$-$10^4$ TeV range, well beyond the reach of the LHC. This work will examine the maximum beam power that can be tolerated for beam energies in the 0.5-8 GeV range. This has implications for how the sensitivity might be further improved with a second generation experiment using an upgraded proton beam from the PIP-II project, which will be capable of providing MW beams to Fermilab experiments later in the next decade.

---

[*] Work supported by Fermi Research Alliance, LLC under contract No. DE-AC02-07CH11359 with the U.S. Department of Energy.

[†] Presented at the Radiation Effects in Superconducting Magnet Materials (RESMM'15) Workshop, FRIB, East Lansing, Michigan, May 2015.

[‡] Email: vspron@fnal.gov

A study of the energy dependence of radiation damage in superconducting coils for a next generation Mu2e at PIP-II

V. Pronskikh, D. Glenzinski, K. Knoepfel, N. Mokhov, R. Tschirhart

Fermi National Accelerator Laboratory

The Mu2e experiment at Fermilab is being designed to study the coherent neutrino-less conversion of a negative muon into an electron in the field of a nucleus. This process has an extremely low probability in the Standard Model, and its observation would provide unambiguous evidence for beyond the standard model physics. The Mu2e design aims to reach a single-event-sensitivity of about 2.5 x $10^{-17}$ and will probe effective new physics mass scales in the $10^3$-$10^4$ TeV range, well beyond the reach of the LHC. This work will examine the maximum beam power that can be tolerated for beam energies in the 0.5-8 GeV range. This has implications for how the sensitivity might be further improved with a second generation experiment using an upgraded proton beam from the PIP-II project, which will be capable of providing MW beams to Fermilab experiments later in the next decade.

The Mu2e experiment at Fermilab [1] will search for evidence of charged lepton flavor violation (cLFV) by observing the conversion of a negative muon into an electron in the Coulomb field of a nucleus without emission of neutrinos. This process is extremely suppressed in the Standard Model, but is predicted to occur at rates observable by Mu2e in a wide variety of new physics models. The Mu2e experiment will probe effective new-physics mass scales in the $10^3$-$10^4$ TeV range. One of the main parts of the Mu2e experimental setup is its target station in which negative pions are generated in interactions of the 8-GeV primary proton beam[§] with a tungsten target, which will be capable of producing ~$2\cdot10^{17}$ negative muons per year. A large-aperture 5-T superconducting production solenoid (PS) enhances pion collection, and an S-shaped transport solenoid (TS) delivers muons and pions to the Mu2e detector. A heat and radiation shield (HRS), installed on the inner bore of the PS, mitigates the effects of radiation dose and heat deposition to protect the PS and the first TS coils from damage. The muons traversing the TS are stopped on an aluminum stopping target situated in the upstream portion of a large aperture detector solenoid (DS), which also houses a tracker and calorimeter in the downstream portion that precisely determine the momenta and energy of particles originating in the stopping target. The Mu2e experiment has a design sensitivity $10^4$ times better than previous muon-to-electron conversion experiments and is scheduled to begin commissioning in 2020 [2].

Regardless of the Mu2e outcome, a next generation experiment, Mu2e-II, with a sensitivity extended another factor of 10 or more, offers a compelling physics case [3]. The improved sensitivity would be enabled by the proposed PIP-II upgrade project, which would significantly improve the Fermilab proton source to enable next-generation intensity frontier experiments [4].

PIP-II is a proposed 250-meter long linac capable of accelerating a 2-mA proton beam to a kinetic energy of 800-MeV corresponding to 1.6-MW of power. Most of the beam will be utilized for the Fermilab Short Baseline Neutrino and Long Baseline Neutrino Facility neutrino programs, but about 200-kW of 800-MeV protons will be available for additional experiments. To achieve another factor of ten improvement in sensitivity, Mu2e-II will require about 100-kW. The linac will have the possibility of being further upgraded to proton energies as high as 3-GeV.

---

[§] In this paper, the term 'beam energy' refers to the kinetic energy of the incoming proton beam.

The present Mu2e design is optimized for 8-kW of protons at 8 GeV. This work uses a MARS15 simulation [5] to study the radiation damage to the PS coils, the peak power deposition in the PS coils, and the stopped muon yield as a function of proton beam energy 0.5 – 8 GeV range. The radiation damage is quantified as displacements-per-atom (DPA). An optimal beam energy would maximize the stopped muon yield while minimizing the DPA damage and heat deposition in the PS coils. For these studies the current Mu2e geometry and beamline are assumed to work for all proton energies so that the simulations begin with the protons interacting on the tungsten production target. This assumption is the focus of a separate study.

The MARS15 Mu2e model is shown in Figure 1. The model includes a description of the superconducting coils in the PS, TS, and DS and the surrounding cryostats as well as collimators situated along the TS for momentum and sign selection. The HRS is situated along the length of the PS inside the cryostat. The tungsten production target is 0.6-cm in diameter and 16-cm long radially centered in the HRS. The aluminum stopping target is modeled as 17 aluminum foils, each 200-μm thick with a ~10 cm diameter, placed 5 cm apart radially, centered in the upstream portion of the DS. For these studies the HRS material was assumed to be either bronze or tungsten. All relevant processes were simulated including production of pions of both signs, their transport, their decay into muons, muon transport and decay, and the stopping of muons in the aluminum target. In the MARS15 simulations, the LAQGSM [6] generator was used for pion and neutron production as well as for other high-energy particle interactions. The MCNP [7] model based on ENDFB-VI [8] was used for neutron transport below 14-MeV.

Details of the PS geometry, including the HRS and production target, are shown in Figure 2. To simulate DPA in the magnet coils the NRT [9] model was used as implemented in MARS15. Below 150 (20) MeV the NRT calculations used the FermiDPA 1.0 [10] cross-section library, which is based on NJOY [11] (ENDFB-VII). Peak DPA and peak power density distributions were calculated in a coil strip located near the beam exit as depicted in Figure 2. The distributions for the total DPA (red line) and the neutron-induced DPA (blue line) are shown in Figure 3. The distributions indicate that the neutron-induced component contributes up to 70% of the total DPA (the rest is photon- and proton-induced).

The limit on DPA was determined based on the requirement that the residual resistivity ratio (RRR), with an initial value of 600, remains at a value of 100 or larger. The calculated dependence of the RRR on the DPA is depicted in Figure 4. The calculations involved experimental uncertainties (blue and green lines) and used the KUR reactor data measured by the COMET collaboration [12]. Figure 4 indicates that the peak DPA damage to the coils should not exceed the $4 \div 6 \cdot 10^{-5}$ before annealing. Experiments show that room temperature annealing of Al-stabilized NbTi can fully recover the RRR. Mu2e plans to anneal the PS coils once a year.

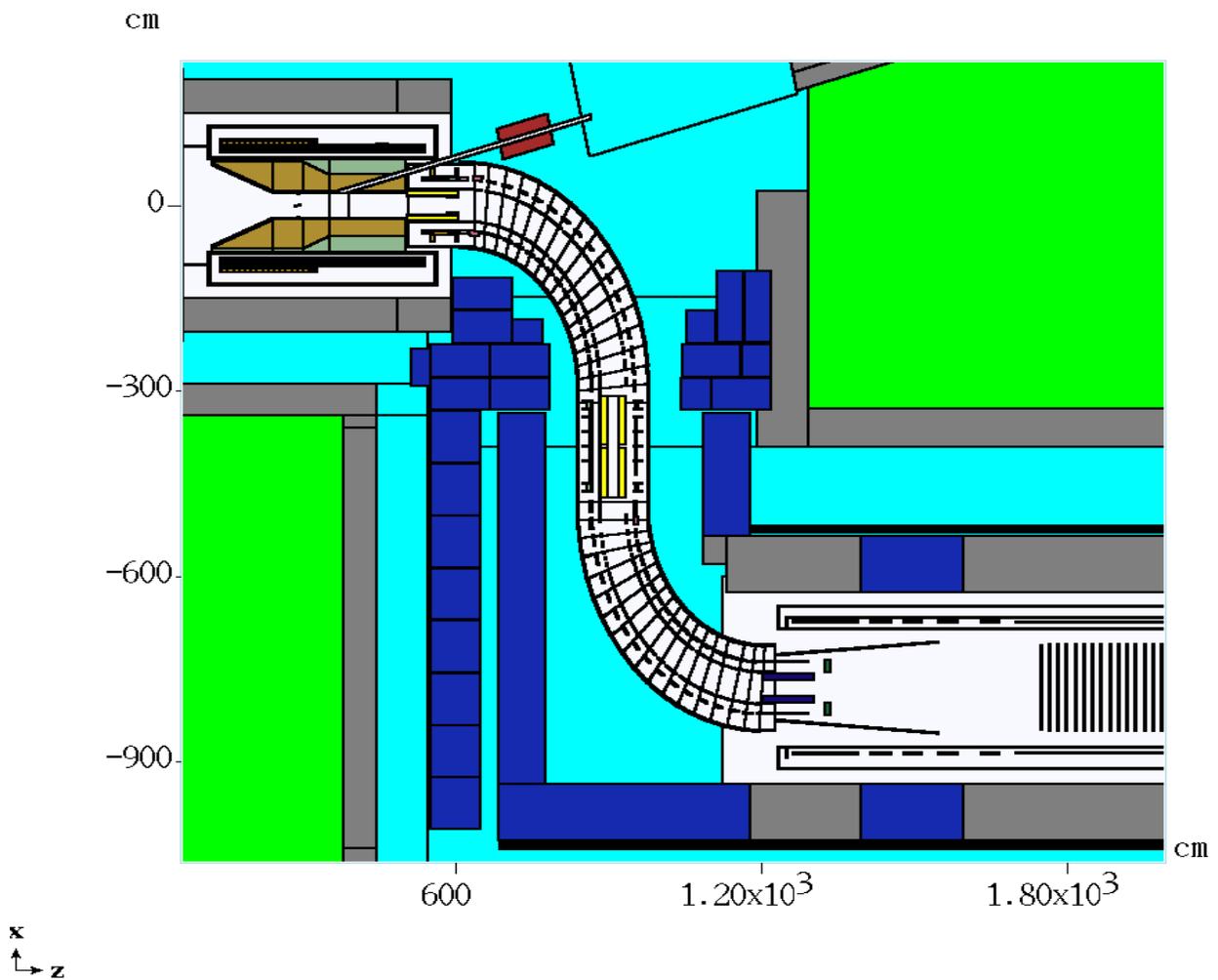

Figure 1. The MARS15 model of Mu2e used in these studies. The PS is located in the top left corner. The incoming proton beam arrives diagonally from right. The S-shaped TS solenoid includes a collimator at each end and at its center for momentum and charge selection. The DS is located in the lower right corner. The aluminum stopping target foils are not shown. At the right end of the DS the first

several tracker stations are shown. The blue and grey blocks represent concrete shielding blocks.

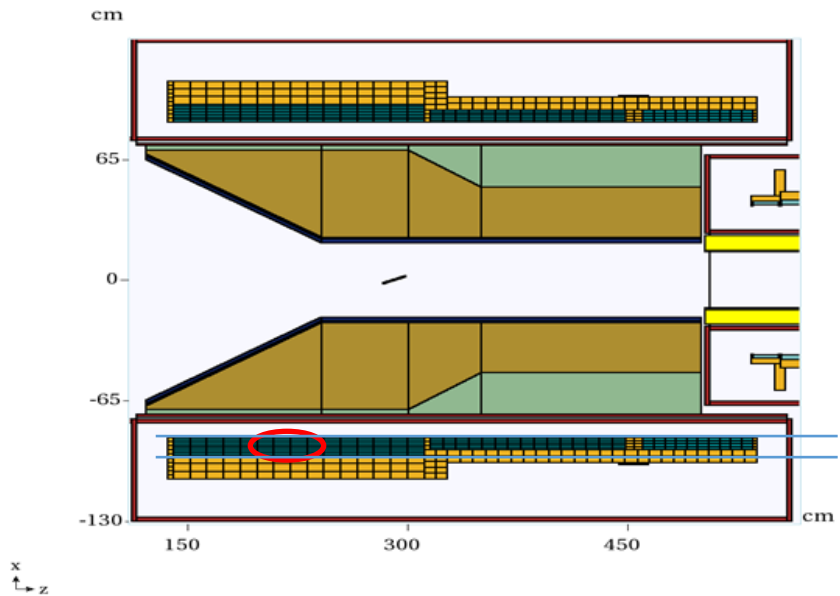

Figure 2. The MARS15 model of the Mu2e PS and HRS. The magnet coils are shown in dark green and yellow while the HRS is shown in brown and light green. The tungsten production target is depicted as the thin diagonal line in the center of the HRS. The incoming proton beam is not shown, but is incident from the top right corner. The first coils of the TS are shown on the right. The region with the largest (peak) DPA and power density is indicated by the red ellipse.

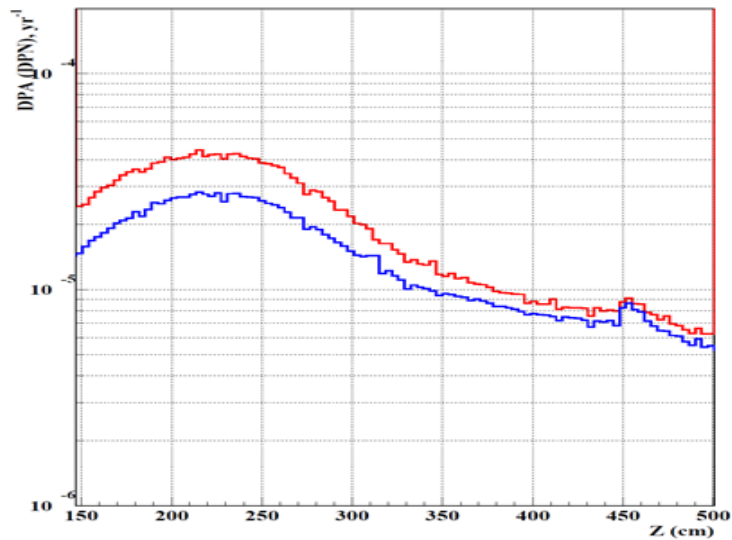

Figure 3. The Total (red line) and neutron-induced (blue line) DPA as a function of length along the PS for 1 year of running at nominal intensity. The start of the TS is located at about z = 475 cm in this plot.

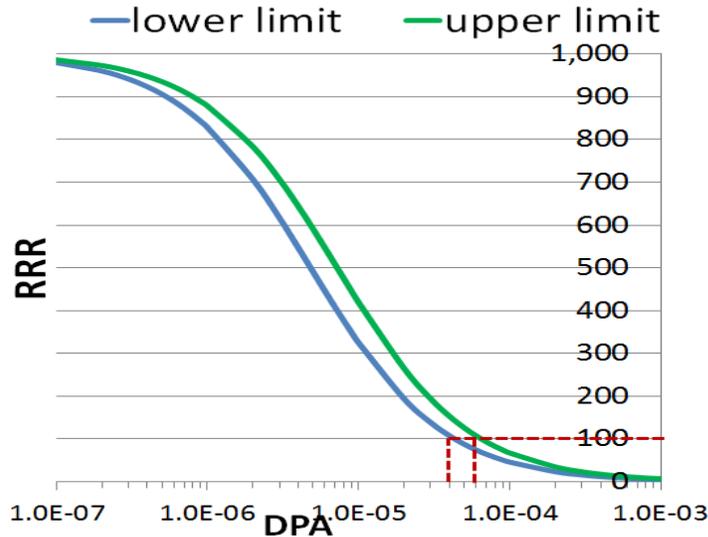

Figure 4. Residual resistivity ratio (RRR) in Al as a function of DPA. The blue and green lines are based on the uncertainties associated with the experimental data used to derive the DPA model.

Figure 3 demonstrates that for the estimated total DPA per year satisfies the requirements defined by Figure 4. The other limits are given in Table 1 [13]. The DPA limit is the most restrictive in the context of this study. The peak absorbed dose for the organic elements of the coils (insulation) is significantly greater than the envisioned lifetime of the experiment and is not a limiting constraint for the sake of the analysis described in this work. It should be noted that the limits for the peak power density and dynamic heat load are sensitive to the design of the magnet-cooling scheme. These limits potentially can be improved with improved cooling.

MARS15 simulations at a variety of proton energies are used to estimate the peak DPA and peak power density as a function of beam energy. The results, per 1 kW of beam power, are shown in Figure 5. A comparison to the current DPA limits indicates that the design with the tungsten HRS can tolerate the 100-kW beam power at a beam energy of 800 MeV, while the peak power density would probably require an improved cooling scheme. Above a proton energy of about 2 GeV the plots are relatively flat with increasing energy.

| Quantity | MARS15 | Limits |
|---|---|---|
| Peak Total Neutron flux in coils, $yr^{-1}$ | $5.2*10^{20}$ | |
| Peak Neutron flux > 100 keV in coils, $yr^{-1}$ | $2.0*10^{20}$ | |
| Peak Power density, µW/g | 17 | 30 |
| Peak DPA, $yr^{-1}$ | $1.4*10^{-5}$ | $4\text{-}6*10^{-5}$ |
| Peak DPA (TS1), $yr^{-1}$ | $1.0*10^{-6}$ | $1.5*10^{-5}$ |
| Peak absorbed dose, MGy/yr | 0.33 | 0.35 |
| Dynamic heat load, W | 28 | 100 |
| Heat load in the HRS volume, kW | 3.3 | |

Table 1. Limits for the main radiation quantities for the Mu2e SC coils, and the expectations for one year of running from the MARS15 simulation.

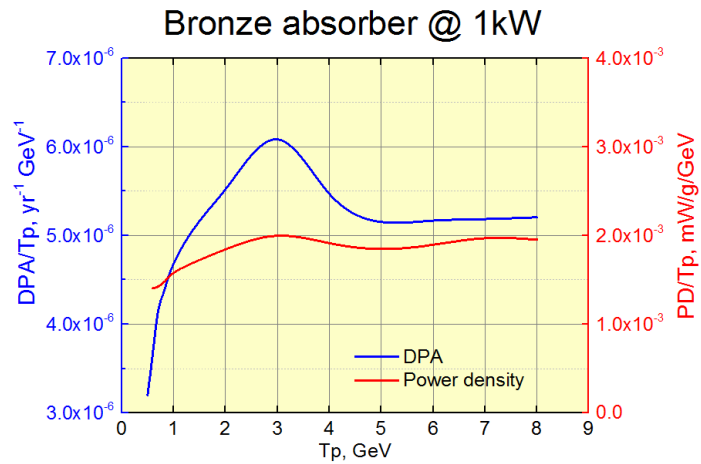

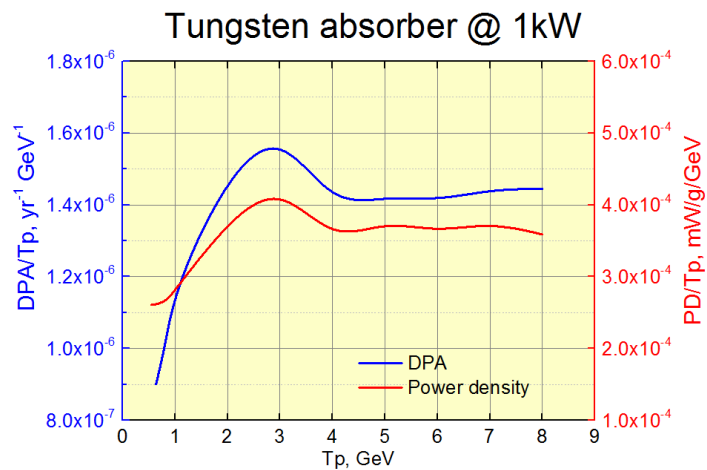

Figure 5. The peak DPA and peak power density per kW for a bronze (top) or tungsten (bottom) HRS as a function of proton beam energy.

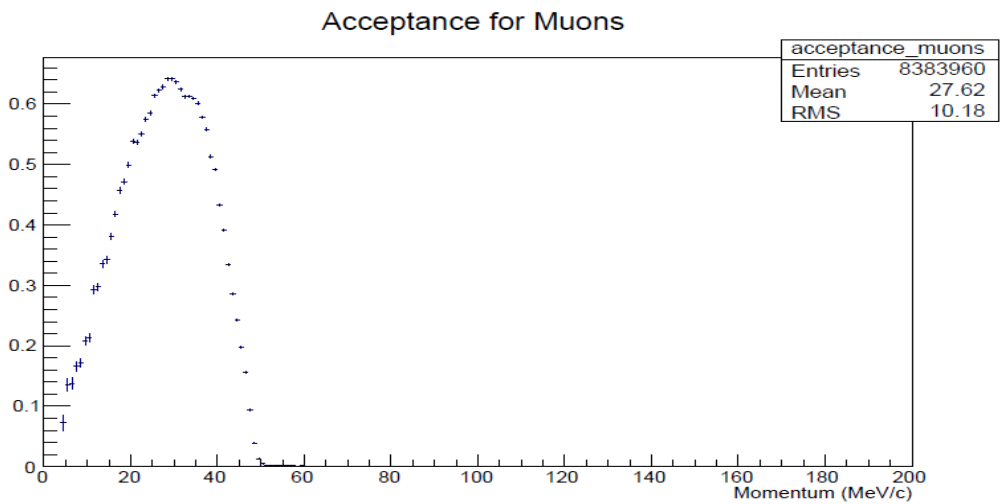

Figure 6. The final acceptance function for muons as described in the text. The horizontal axis corresponds to the muon momentum as it enters the TS while the vertical axis corresponds to the probability of yielding a stopped muon at the aluminum target in the DS.

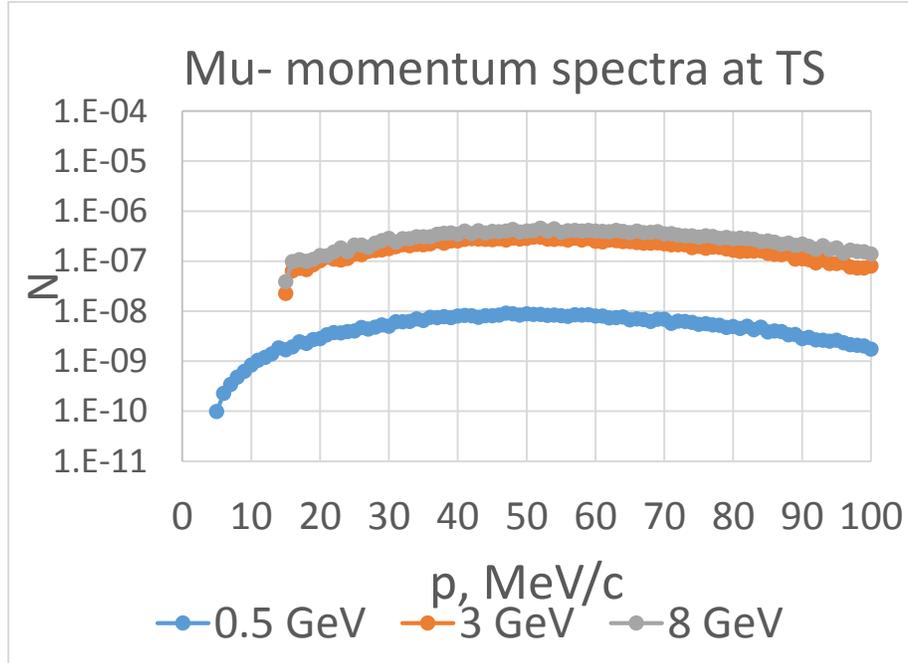

Figure 7. The momentum spectra of muons entering the TS from the MARS15 model for a proton beam energy of 0.5, 3, and 8 GeV normalized per proton-on-target.

For this work a figure-of-merit was defined as the ratio of the number of stopped muons to the peak DPA for a constant beam power at a given proton energy. The peak DPA was taken from Figure 5 while the number of stopped muons is calculated as described below.

The number of stopped muons was estimated by convoluting the momentum spectra of muons and pions arriving at the TS entrance with an acceptance function corresponding to the probability of a muon stop as a function of particle momentum. A G4Beamline simulation [14], which provides a fast and accurate beam-transport model, was used to derive acceptance functions separately for muons and pions. Samples, each corresponding to $\mathcal{O}(10^8)$ protons-on-target, were generated at a variety of proton energies. The muon (pion) related acceptance function was derived by calculating the probability that a muon (pion) of a given momentum as it entered the TS would yield a stopped muon in the aluminum target in the DS. The resulting acceptance functions were consistent across beam energies for muons and pions separately. Consequently, to minimize the associated statistical uncertainties, acceptance functions formed by combining the simulation samples from all beam energies were utilized. The resulting muon acceptance function is shown in Figure 6. This is convoluted with the muon momentum spectrum at the entrance to the TS as estimated using the MARS15 simulations at varying proton energies. Some sample muon spectra are shown in Figure 7 while the number of muons entering the TS (per proton on target) as a function of proton energy is shown in Figure 8. The total number of stopped muons is determined by summing the contribution from the muons and pions entering the TS. The resulting total number of stopped muons for a 3 year run (6 x $10^7$ seconds of run time) at 100 kW is shown in Figure 9 as a function of proton energy.

The stopped muon yields are combined with the peak DPA estimates discussed above to obtain the figure-of-merit as a function of proton energy as depicted in Figure 10. By this metric the optimal beam energy lay in the 1-3 GeV range and varies slowly above about 4 GeV. The figure-of-merit is about the same at 800 MeV as it is at 8 GeV proton beam energy.

The single-event-sensitivity as a function of proton beam energy can be estimated from Figure 8 for a nominal 3 year run at 100 kW of beam. This is shown in Figure 11. It is assumed that the limitations discussed above can be mitigated, that the detector performance remains unchanged at the higher beam intensities, and that an aluminum stopping target is used. This may be considered an optimistic estimate.

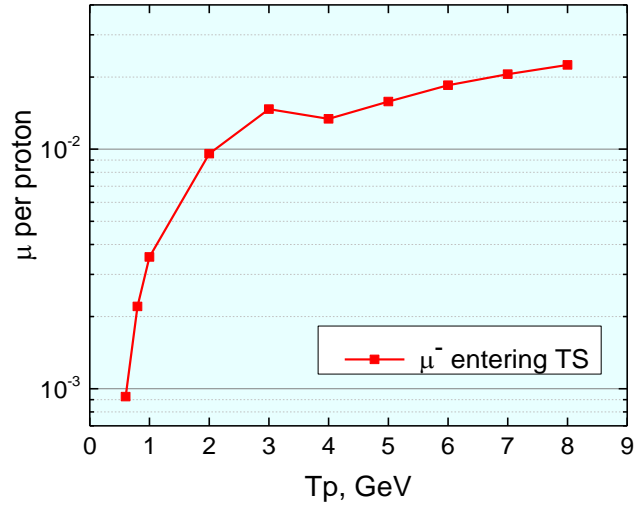

Figure 8. The number of muons entering the TS per proton-on-target as a function of proton beam energy as estimated in the MARS15 model. The LAQGSM production generator used by MARS transitions from one production model to another in the 3-4 GeV region, which gives rise to the observed kink at a proton beam energy of 3 GeV. Tp is the proton kinetic energy.

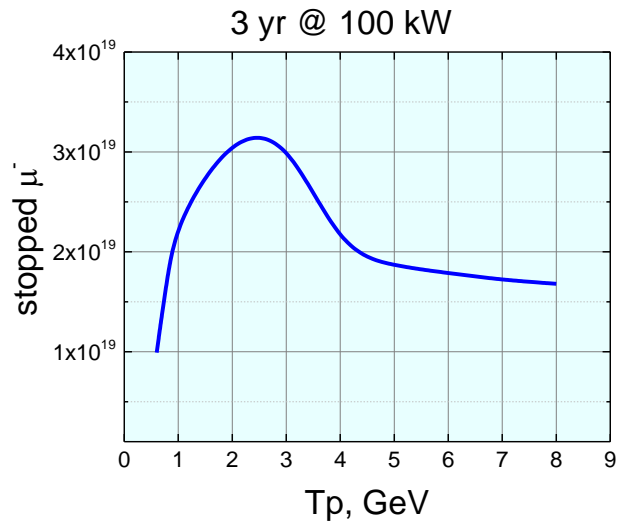

Figure 9. The number of stopped muons as a function of proton energy for a nominal 3 year run with 100 kW of beam. Tp is the proton kinetic energy.

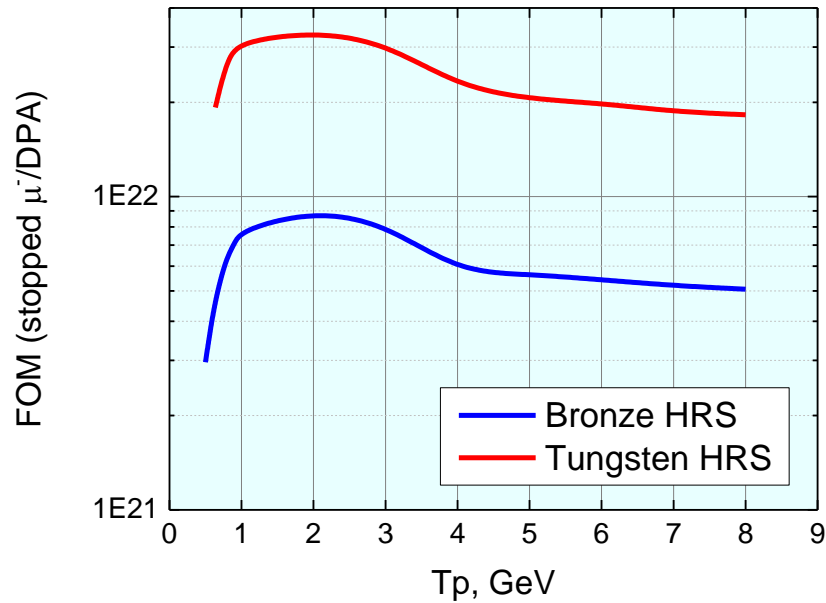

Figure 10. The figure-of-merit (FOM) as a function of proton beam energy calculated as described in the text. Tp is the proton kinetic energy.

The estimate of the single-event-sensitivity can be repeated assuming instead that the DPA limitations discussed above cannot be mitigated and thus limit the maximum beam power that can be tolerated at a given proton beam energy. This will reduce the total number of stopped muons that can be produced for a given run time and thus degrade sensitivity. The resulting estimates, for a nominal 3 year run at the maximum beam power allowed assuming DPA limitations, as a function of proton beam energy are shown in Figure 12. This may be considered a pessimistic scenario.

**Conclusions**

The radiation damage to the superconducting coils of the Mu2e production solenoid (quantified as displacements per atom, DPA), the power deposition limitations on the performance of the Mu2e production solenoid, and the muon-stopping rate are estimated for a variety of potential beam-upgrade scenarios that would enable a next generation Mu2e-II experiment. Utilizing a MARS15 simulation model that includes a detailed description of the current Mu2e beamline, shielding, and detector geometry the above quantities are estimated for proton beam energies in the range 0.5 – 8 GeV for beam intensities up to 100 kW. A heat and radiation shield (HRS) located inside the production solenoid mitigates the radiation damage and power deposition effects on the superconducting coils. In these studies the HRS was assumed to be composed of either bronze or tungsten. The DPA, power deposition and muon-stopping rate all peak at a proton beam energy of 2-3 GeV.

Taking at face value the current estimates of the maximum DPA the PS coils can tolerate before performance is affected, these MARS15 simulations suggest that the current production solenoid with the current HRS geometry could tolerate beam power of <100 kW in the 0.8 – 8 GeV range of beam energies.

An ad hoc figure-of-merit is constructed from the ratio of stopped muons to DPA in the production solenoid coils and suggests that the optimal proton beam energy would be in the 1-3 GeV range. The figure-of-merit is up to about 30% worse outside of this optimal range over the range of energies considered. It is about the same at a beam energy of 0.8 GeV as it is at 8 GeV.

A single-event-sensitivity is estimated for Mu2e-II, as described in the text, as a function of proton beam energy assuming the beam power is limited by the DPA damage to the PS coils and its detrimental effects to the superconductor residual resistivity ratio. Under this somewhat pessimistic scenario the sensitivity can be improved by as much as a factor of about 5, relative to the current Mu2e estimate, for a nominal 3 year run utilizing an aluminum stopping target (cf. Figure 12). More optimistically, if the DPA effects can be mitigated to allow running at 100 kW, the single-event-sensitivity can improve by a factor of 10 or more for a nominal 3 year run utilizing an aluminum stopping target (cf. Figure 11).

Strategies to extend the DPA limitations of the production solenoid are under study and include, for example, the possibility of more frequent anneals, an improved HRS geometry, or an improved production solenoid. Other parts of the Mu2e apparatus may also have limitations that would imply a beam intensity of < 100 kW. These are the subject of separate on-going studies.

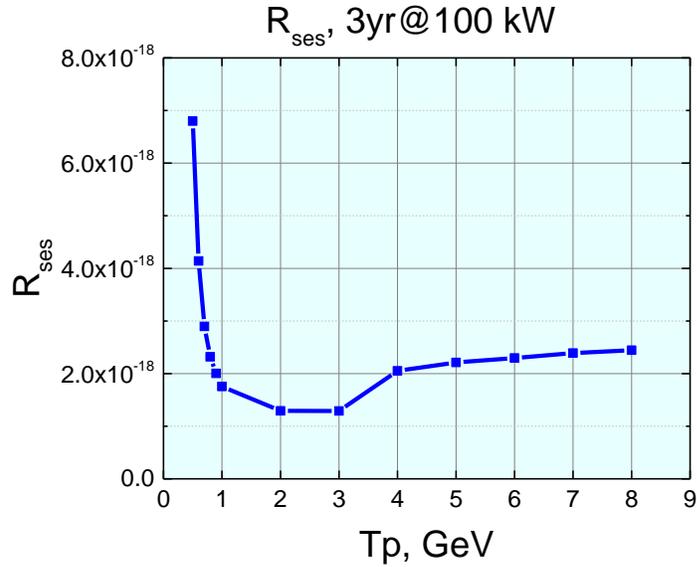

Figure 11. An estimate of the single-event-sensitivity (SES) for a nominal 3-year run at 100 kW for various proton beam energies (Tp). The assumptions affecting this estimate are discussed in the text. By comparing to the estimated Mu2e SES (2.5 x 10⁻¹⁷ for 3 years of 8 kW of proton beam at 8 GeV), this suggests that a next-generation Mu2e-II experiment might plausibly improve the sensitivity by a factor of 10 utilizing an upgraded proton source. The assumptions here might be considered optimistic.

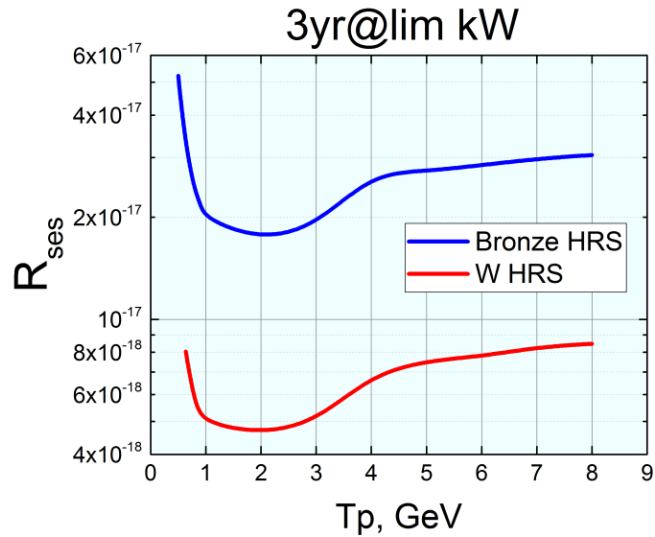

Figure 12. A pessimistic estimate of the single-event-sensitivity for a nominal 3-year run for various proton beam energies (Tp). The calculation here is the same as used in Figure 11 except that here it is assumed that the DPA limitations discussed earlier cannot be mitigated and thus limit the maximum beam power to < 100 kW.